\documentclass[
  twocolumn,
  aps,prb,
  superscriptaddress,
  showpacs,
  amsmath,amssymb,
  longbibliography
]{revtex4-2}

\usepackage{epsf}

\usepackage{natbib}
\setcitestyle{square,numbers}

\usepackage{tabularx} 
\usepackage{graphicx} 
\usepackage{epstopdf}
\usepackage{epsf}
\usepackage{dcolumn}
\usepackage{bm}
\usepackage{color, xcolor}

\usepackage{multirow}

\usepackage[normalem]{ulem}

\usepackage{color,soul}
\setul{0.5ex}{0.5ex}
\setulcolor{red}

\begin{document}

\title{Exchange interaction in ACu$_3$Fe$_2$Re$_2$O$_{12}$ quadruple perovskites}

\author{Fedor Temnikov}
\affiliation{Institute of Metal Physics, S. Kovalevskaya str. 18, 620108 Ekaterinburg, Russia}

\author{Alexey V. Ushakov}
\affiliation{Institute of Metal Physics, S. Kovalevskaya str. 18, 620108 Ekaterinburg, Russia}

\author{Evgenia V. Komleva}
\affiliation{Institute of Metal Physics, S. Kovalevskaya str. 18, 620108 Ekaterinburg, Russia}
\author{Zhehong Liu}
\affiliation{Beijing National Laboratory for Condensed Matter Physics, Institute of Physics, Chinese Academy of Sciences, Beijing 100190, China}

\author{Youwen Long}
\affiliation{Beijing National Laboratory for Condensed Matter Physics, Institute of Physics, Chinese Academy of Sciences, Beijing 100190, China}

\author{Valentin Yu. Irkhin}
\affiliation{Institute of Metal Physics, S. Kovalevskaya str. 18, 620108 Ekaterinburg, Russia}

\author{Sergey V. Streltsov}
\affiliation{Institute of Metal Physics, S. Kovalevskaya str. 18, 620108 Ekaterinburg, Russia}

\date{\today}

\begin{abstract}
Quadruple perovskites ACu$_3$Fe$_2$Re$_2$O$_{12}$ attract considerable interest due to their high Curie temperatures (up to $710$K), which strongly depend on the A-site cation. 
In this work, we employ first-principles calculations to investigate their electronic structure and magnetic exchange interactions.  A band mechanism of magnetism that explains the antiferromagnetic character of the exchange interactions and their strong dependence on the filling of the Re $t_{2g}$ states is proposed. These antiferromagnetic interactions stabilize ferrimagnetic ground state. The calculated Curie temperatures, obtained within the Onsager reaction field theory, are in a good agreement with experimental data.

\end{abstract}

\maketitle

\section{Introduction}

\begin{figure}[b!]
\includegraphics[width=0.46\textwidth]{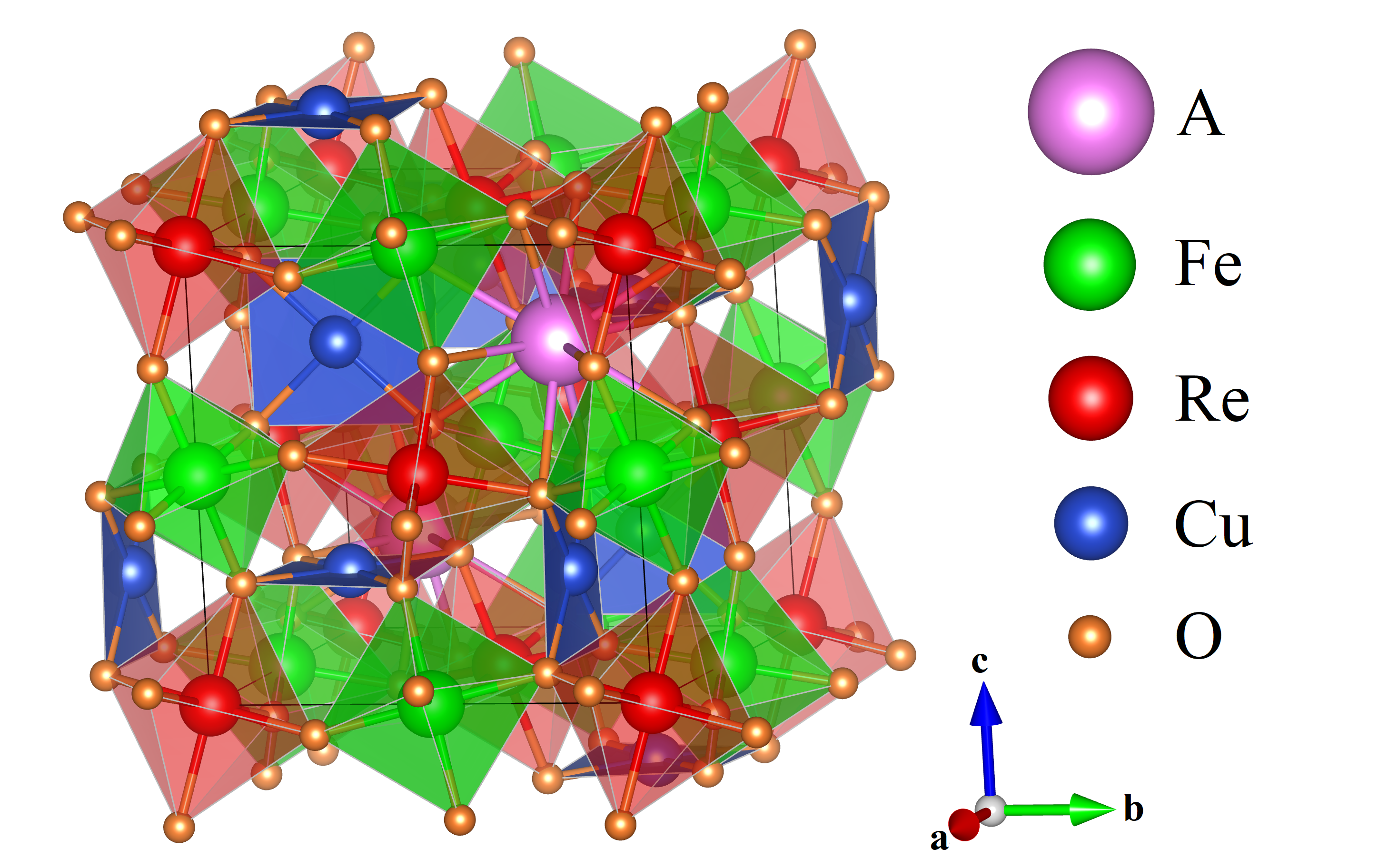}
\caption{\label{fig:crystal-structure} The crystal structure of ACu$_3$Fe$_2$Re$_2$O$_{12}$ quadruple perovskites.}
\end{figure}

As a natural generalization of ABO$_3$ perovskite structures, quadruple perovskites with the chemical formula AA$'$$_3$B$_2$B$'$$_2$O$_{12}$ represent a rich platform that allows for variation in both A- and B-site ions to obtain materials with extraordinary physical properties.   In these compounds, the A-site cation is typically a rare-earth, alkali, or alkaline-earth metal with a relatively large ionic radius, while the A$'$ site hosts a Jahn-Teller active transition metal ion, forming square A$'$O$_4$ plaquettes. The B and B$'$ sites, which can be occupied by identical or different transition metal ions, are coordinated by oxygen atoms in an octahedral geometry, see Fig.~\ref{fig:crystal-structure}. Since A$'$-, B-, and B$'$-site ions can all be magnetic, quadruple perovskites may exhibit complex magnetic structures.

Similar to ``conventional''  ABO$_3$ perovskites, quadruple perovskites show a variety of different physical phenomena. For example, (Ca/La/Bi)Cu$_3$Mn$_4$O$_{12}$ exhibit significant low-field magnetoresistance \cite{magnetoresistance1,magnetoresistance2,magnetoresistance3}. CaCu$_3$Ti$_4$O$_{12}$ is known for its extremely high dielectric constant with weak temperature dependence \cite{CaCuTiO} and has also been proposed as a potential piezoelectric material in the presence of copper vacancies \cite{CaCuPiezo}. Ferrimagnetism with relatively high critical temperatures is observed in a series of RCu$_3$Mn$_4$O$_{12}$, R $=$ rare-earth metal, ($T_C$ up to 400~K) \cite{RCuMnO, RCuMnO2}, as well as in CaCu$_3$Cr$_2$Re$_2$O$_{12}$ ($T_C=360$~K)  \cite{CaCuCrReO}.

Another class of quadruple perovskites characterized by high ferrimagnetic transition temperatures is ACu$_3$Fe$_2$Re$_2$O$_{12}$.
$T_C$ in these compounds varies over a wide range  of 170-710 K \cite{Chen2014, liu2022, Liu2024, Zhang2025} depending on the A cation. Moreover, first-principles calculations have predicted half-metallic ferrimagnetism in several members of this series, with an energy gap of approximately 2~eV in the non-conducting spin channel \cite{Liu2024, Zhang2025, poteryaev2025}. This makes ACu$_3$Fe$_2$Re$_2$O$_{12}$ an extremely interesting candidate for spin filters, whose efficiency depends on three key parameters: the spin-majority channel gap (which blocks thermally activated conductance), the magnetization (determining spin polarization), and the Curie temperature (defining the operational temperature range)~\cite{liu2022}.

While an impressive progress has been recently achieved in synthesis of such materials as LaCu$_3$Fe$_2$Re$_2$O$_{12}$ and DyCu$_3$Fe$_2$Re$_2$O$_{12}$ with $T_C= 710$~K and 650~K, respectively, theoretical understanding is still lacking. In addition to calculations for a few quadruple perovskites~\cite{Liu2024,Zhang2025} a possible role of the number of $d$-electrons on Re was suggested~\cite{wang2021}. 

In this paper we studied exchange interaction in ACu$_3$Fe$_2$Re$_2$O$_{12}$ by means of {\it ab initio} calculations. It is shown that these are Cu-Re and Fe-Re exchange interactions which determine magnetic ordering. Interestingly, a naive superexchange model does not seem to explain details of exchange coupling, so that we propose a model based on band magnetism. This model naturally explains increase of Curie temperature with the number of Re $t_{2g}$ electrons. The Onsager reaction field theory, which goes beyond the mean-field approximation, reproduces this trend observed experimentally.

\section{Calculation details}

First-principles density functional theory (DFT) calculations were performed using the Vienna Ab initio Simulation Package (VASP) \cite{kresse1996} in the generalized gradient approximation \cite{perdew1996}. To account for strong on-site electron correlations in the transition metal $d$ states, the GGA+$U$ method in the Dudarev formulation \cite{dudarev1998} was applied, using the following values of $U-J_H$:  7~eV for Cu \cite{sun2022}, 4.1~eV for Fe \cite{hong2012}, 1.5~eV for Re \cite{lim2016}. 

The energy cutoff is set to 500~eV and the reciprocal space division is 8$\times$8$\times$8 k-points. The experimental crystal structure data for all ACu$_3$Fe$_2$Re$_2$O$_{12}$ compounds with the space group Pn\=3 (No. 201) reported in \cite{Chen2014} (A=Ca), \cite{liu2022} (A=La), \cite{Liu2024} (A=Dy), \cite{Zhang2025} (A=Na) and  (A=Cu, Ag, Ce) \cite{Long}, were used. 

Although the electron states in different bands have different degree of localization, a simple effective Heisenberg model 
\begin{eqnarray}
\label{eq:Heisenberg}
H = \sum_{i>j} J_{ij} {\bf S}_i {\bf S}_j,
\end{eqnarray}
is as a reasonable approximation. Here $i$ and $j$ numerate sites and summation runs once over each index pair. Spins for Cu$^{2+}$ ($S=1/2$) and Fe$^{3+}$ ($S=5/2$) are chosen according to their valencies, while for Re we fix $S=1$ to be able to analyze properties of all materials within the same model. The exchange constants $J_{ij}$ were obtained by Wannier function projection technique using the Green's function approach \cite{korotin2015}.

\begin{figure}[t!]
\includegraphics[width=0.5\textwidth]{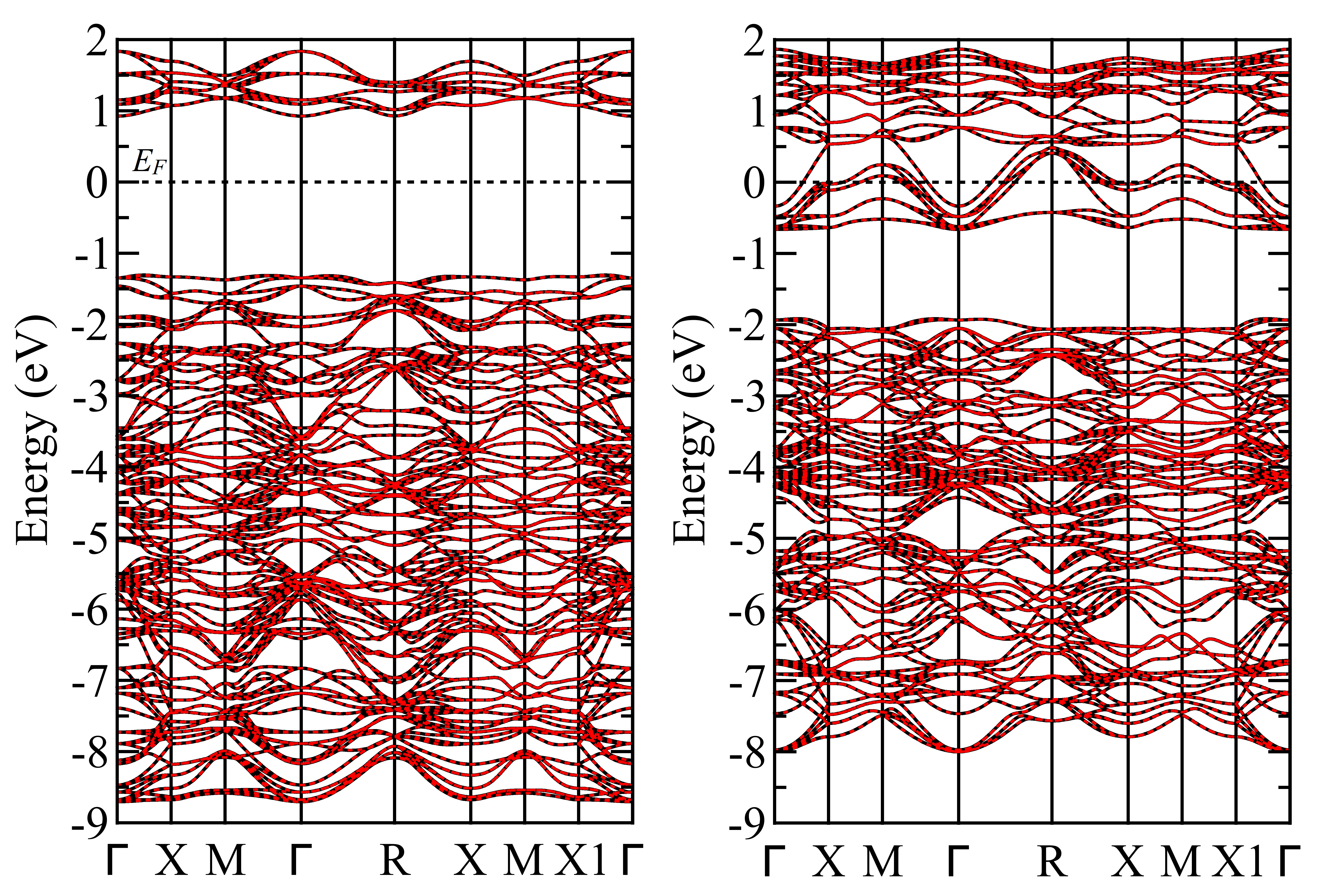}
\caption{\label{fig:Wannierbands} Band structures comparison of NaCu$_3$Fe$_2$Re$_2$O$_{12}$ obtained by DFT+U calculation. Initial DFT+U bands are shown in black, projected on Wannier functions are in red. The left panel illustrates the majority-spin channel, and the right panel -- the minority-spin channel.}
\end{figure}

\begin{figure}[t!]
\includegraphics[width=0.5\textwidth]{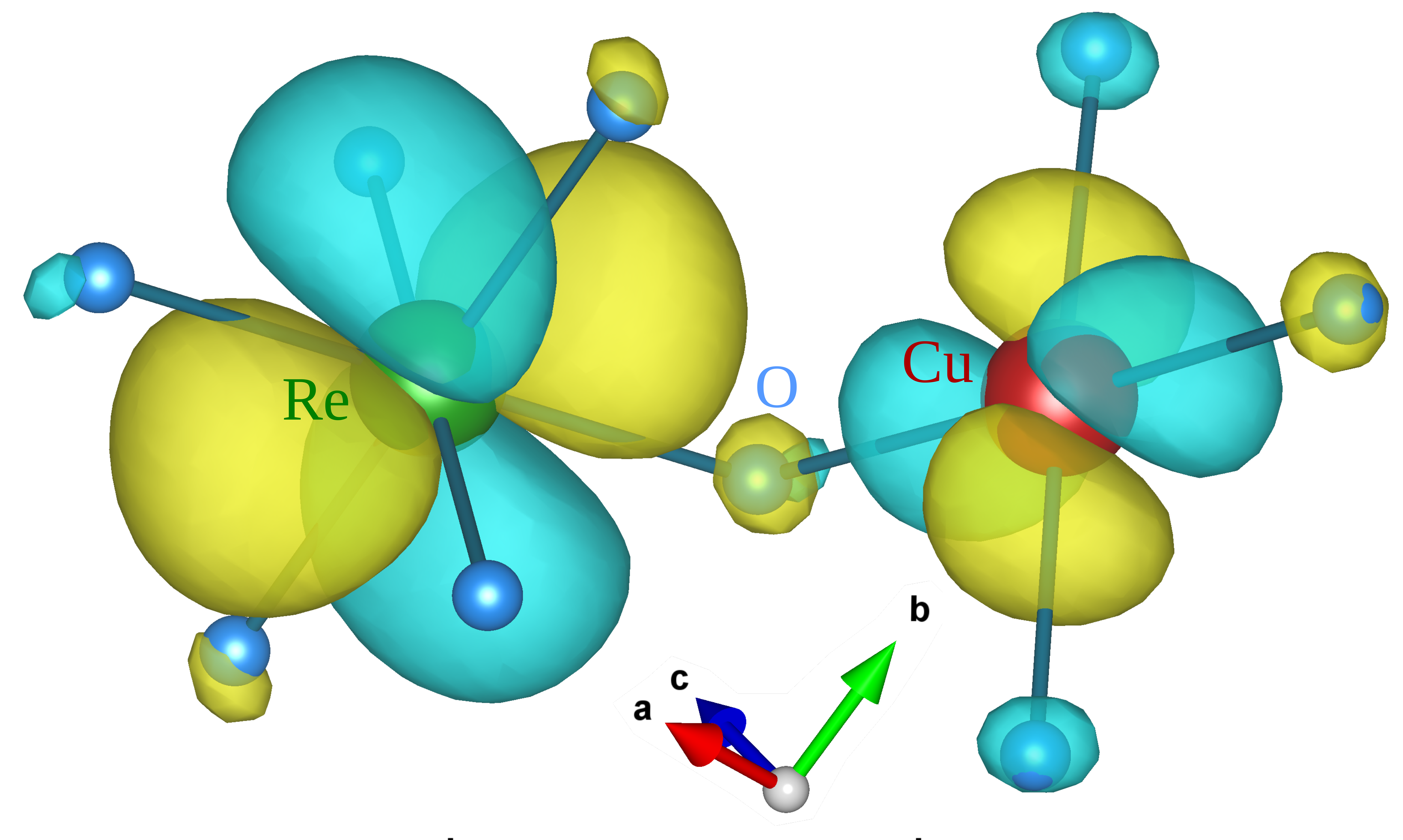}
\caption{\label{fig:Wannierorbitals} Wannier orbitals (Cu $x^2-y^2$ and Re $xy$) employed for the calculation of the exchange parameter $J_{\rm Cu-Re}$.}
\end{figure}

The correctness of the Wannier projection is confirmed by the agreement between the DFT+U band structure and that reconstructed from the Wannier functions; the NaCu$_3$Fe$_2$Re$_2$O$_{12}$ case is shown in Fig.~\ref{fig:Wannierbands}. The projection was performed onto the subspace of O-2$p$, Cu-3$d$, Fe-3$d$, and Re-$t_{2g}$ states. A visualization of the Wannier orbitals corresponding to the exchange parameter $J_{Cu-Re}$ is presented in Fig.~\ref{fig:Wannierorbitals}.

\begin{figure}[t!]
\includegraphics[width=0.48\textwidth]{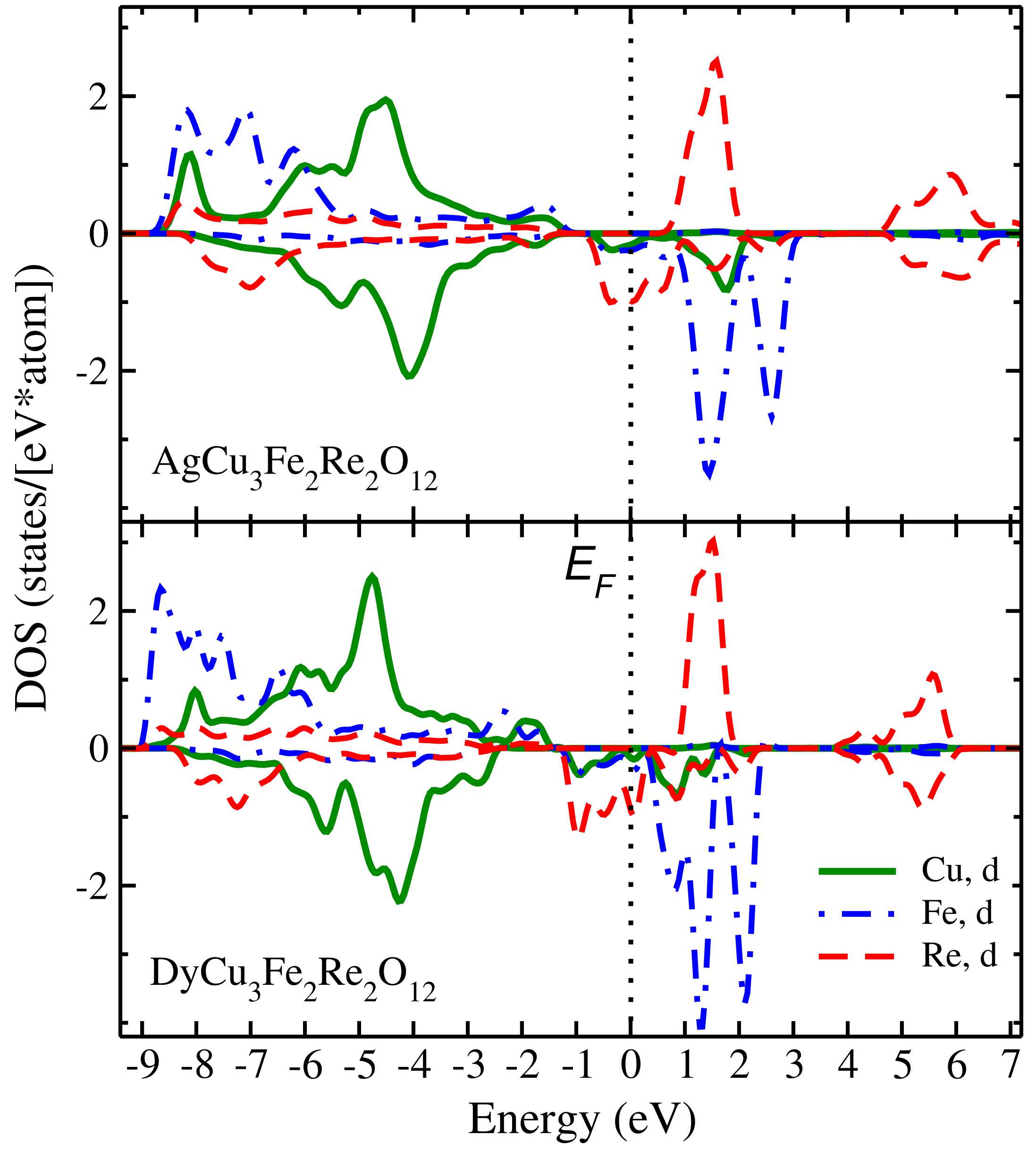}
\caption{\label{fig:DOS} Density of states (DOS) as obtained in DFT+U calculations. Due to different valences of A ions Re  is expected to have $ n_{{\rm Re-d}}^{{\rm nominal}}=1.5$ for Ag or $ n_{{\rm Re-d}}^{{\rm nominal}}=2.5$ electrons on $t_{2g}$ shell for Dy. Positive (negative) DOS correspond to spin majority (minority).}
\end{figure}

\section{DFT+U results }

\begin{table}[t!]
\centering
\begin{tabular}{cccccccccc}
\hline
\hline
       A                 & Cu   & Ag  & Na  & Ca  & Dy   & La   & Ce \\
\hline
$m_{\rm s}$(Cu), $\mu_B$ &  0.6 & 0.6 & 0.5 & 0.6 &  0.5 & 0.5  & 0.5 \\
$m_{\rm s}$(Fe), $\mu_B$ &  4.1 & 4.1 & 4.2 & 4.1 & 4.1 & 4.0 & 4.1  \\
$m_{\rm s}$(Re), $\mu_B$ & -0.8 & -0.8 & -0.7 & -1.1 & -1.3 & -1.3 & -1.4 \\
$ n_{{\rm Re-d}}^{{\rm nominal}}$ & 1.5 & 1.5 & 1.5 & 2 & 2.5 & 2.5 & 2.5 \\
\hline
$m_{\rm total}^{\rm calc}$, $\mu_B$ & 10.0  &  10.0 & 10.0  & 8.0  & 16.3*  &  8.0 & 8.0 \\
$m_{\rm total}^{\rm exp}$, $\mu_B$ &  $\sim$5 &  --  & 9.4  &  8.7 &  14.0** & 8.0  & 8.0 \\
\hline
\hline
\end{tabular}
\caption{\label{Tab:Moments} Calculated  spin moments obtained by DFT+U method and experimental saturation magnetic moments for ACu$_3$Fe$_2$Re$_2$O$_{12}$ with different A. 
*Includes contribution from the $A$-site ion. 
**In the field of 7~T, the total magnetic moment is not saturated.
}
\end{table}

\subsection{Electronic structure and origin of half-metallicity}

We start by describing the electronic structure of $\text{ACu}_3\text{Fe}_2\text{Re}_2\text{O}_{12}$, as it critically influences the magnetic properties discussed later. All  these materials are half-metals in DFT+U calculations. Fig.~\ref{fig:DOS} shows the density of states (DOS) for two representative quadruple perovskites with different number of Re $t_{2g}$ electrons.

Strong Hubbard repulsion on Fe, Cu, and the rare-earth element (if present) shifts the $3d$ (or $4f$) states away from the Fermi level.  Re $5d$ (in particular, $t_{2g}$) bands are much less correlated ($U \sim 1.5 - 2\,\text{eV}$ is of order of the $t_{2g}$ bandwidth ~\cite{kim2016,lim2016, scudder2021}) and more hybridized with O $2p$. Consequently, the system exhibits no Mott gap or well-defined Hubbard bands in DFT+U. Instead, we observe a Stoner-like splitting, with the Fermi level crossing the spin-minority Re $t_{2g}$ band while a gap forms in the spin-majority channel (Re $e_g$ states are at $\sim$ 5-6 eV). The strong antiferromagnetic (AFM) Fe-Re and Re-Cu exchange interactions, whose mechanism will be discussed below, ensure that all Re ions are ferromagnetically aligned. As a result  ACu$_3$Fe$_2$Re$_2$O$_{12}$ materials turn out to be half-metallic on DFT+U(+SOC) level, while dynamical correlation effect can potentially modify this situation~\cite{poteryaev2025}. 

\begin{figure}[t!]
\includegraphics[width=0.5\textwidth]{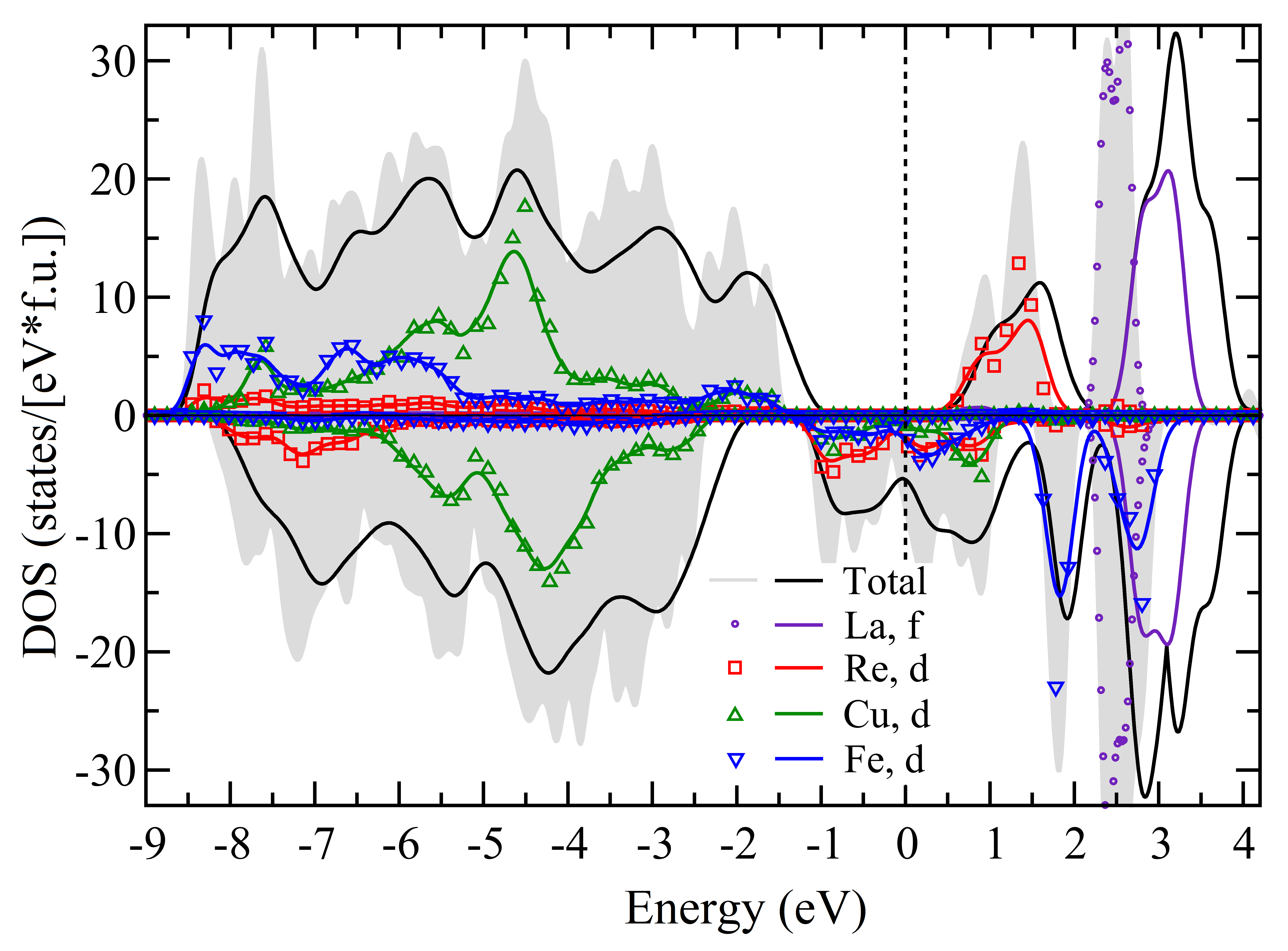}
\caption{\label{fig:SOC_DOS} Density of states (DOS) as obtained in DFT+U (symbols for partial DOS and shaded background for total DOS) and DFT+U+SOC (solid lines) calculations for LaCu$_3$Fe$_2$Re$_2$O$_{12}$. }
\end{figure}

Spin-orbit coupling (SOC) playing a minor role in quadruple perovskites despite the fact that Re is 5$d$ element. SOC leads to a small canting of the Fe and Re magnetic moments~\cite{Liu2024}. A comparison of the density of states obtained by DFT+U and DFT+U+SOC calculations is presented in Fig.~\ref{fig:SOC_DOS} for one of the materials -- LaCu$_3$Fe$_2$Re$_2$O$_{12}$. Inclusion of spin-orbit coupling (SOC) leads to a reduction of the band gap in the spin-majority channel by $\sim0.4$~eV. A noticeable suppression of DOS at the Fermi level is  observed. In addition, the La $f$-states are shifted toward higher energies.

\subsection{Magnetic moments}
Spin moments obtained by DFT+U calculations are summarized in Table~\ref{Tab:Moments}. The magnetic moments on Cu and Fe ions are nearly independent of the A-site ion, indicating that their magnetic states remain unchanged. The moment of the rhenium ion increases with the nominal number of electrons. On average, $m_{\rm s}$(Re)  is approximately half of the nominal theoretical value, suggesting both a strong hybridization with ligand $2p$ states and possible itinerant character of rhenium electrons. 

The total magnetic moments per formula unit obtained from DFT$+U$ calculations are in excellent agreement with the experimental saturation moments for A=La and A=Ce, and small deviations are observed for A=Na and A=Ca. In the case of A=Dy, the theoretical value exceeds the experimental one due to the fact that the full saturation has not been achieved experimentally \cite{Liu2024}. A significant difference, approximately by a factor of two, is observed for the case of CuCu$_3$Fe$_2$Re$_2$O$_{12}$. As shown in Ref.~\cite{poteryaev2025}, taking into account strong correlations within the DFT+DMFT approach improves the agreement with experiment, reducing the total magnetic moment to $m_{\text{total}}^{\text{calc}} = 7.6~\mu_{\text{B}}$. However, the true origin of this discrepancy remains unclear at the moment and further experimental (including possible issues with disorder) and theoretical studies have to be done for this material.

\subsection{Exchange parameters}
\begin{table}[b!]
\centering
\begin{tabular}{lccccccccc}
\hline
\hline
                  & Cu & Ag & Na & Ca & Dy & La & Ce \\
\hline
$J_{{\rm Fe - Re}}$, K & 72 & 73 & 40 & 103 & 84 & 73 & 78  \\
$J_{{\rm Fe - Cu }}$, K & 41  & 41  & 23 &  42 &  41 & 24  & 41 \\
$J_{{\rm Cu - Re}}$, K &  364 & 356 & 314 & 506 & 530 & 496 & 528  \\
\hline
$ n_{{\rm Re-d}}^{{\rm nominal}}$ & 1.5 & 1.5 & 1.5 & 2 & 2.5 & 2.5 & 2.5 \\
\hline
$ T_C^{exp}$, K & 200 & 174 & 240 & 560 & 650 & 710 & 625 \\

\hline

$ T_C^{\rm MF}$, K & 865 & 858 & 675 & 1246 & 1197 & 1122 & 1170 \\

$ T_C^{\rm ORF} $, K & 444 & 441 & 341 & 641 & 609 & 569 & 594 \\

\hline
\hline
\end{tabular}
\caption{\label{Tab:FeRe-exchanges} Results of DFT+U calculations for ACu$_3$Fe$_2$Re$_2$O$_{12}$ with different A and experimental  $T_C$. The Heisenberg model is defined in \eqref{eq:Heisenberg} and for Fe$^{3+}$ ($3d^5$) we choose $S=5/2$, for Cu$^{2+}$ ($3d^9$) -- $S=1/2$; while the number of electrons on Re, $ n_{{\rm Re-d}}^{{\rm nominal}}$, changes in this series for Re, we fix $S=1$ for convenience. MF stands for mean-field (estimation of Curie temperature), and ORF for Onsager reaction field theory (see below Sect. IV).
} 
\end{table}

Results of the exchange interaction of various quadruple perovskites with general chemical formula ACu$_3$Fe$_2$Re$_2$O$_{12}$ are summarized in Table ~\ref{Tab:FeRe-exchanges}. One can see that all materials exhibit the same trend: the strongest interaction is formally the antiferromagnetic (AFM) Cu–Re exchange 
$J_{\rm {Cu-Re}}$, which is at least five times larger than the second-strongest interaction, the AFM Fe–Re coupling $J_{\rm {Fe-Re}}$. In fact, these interactions contribute nearly equally to the magnetic energies, if one takes into account different spins of Cu and Fe ions (e.g., energy per Cu-Re bond is 182~K, while it is 180~K per Fe-Re bond in the case of A=Cu). This agrees with results of previous theoretical study~\cite{wang2021}.

The last important exchange coupling is between Fe and Cu ions, it is AFM, but much smaller than $J_{\rm {Cu-Re}}$ and $J_{\rm {Fe-Re}}$ and cannot override them. As one case see from Fig.~\ref{fig:crystal-structure}, these three ions form a Fe-Re-Cu triangles. Therefore, two large AFM exchanges make spins on remaining bond, Fe-Cu, be FM.

One clear tendency evident from Table~\ref{Tab:FeRe-exchanges} is the increase in both Cu-Re and Fe-Re exchange interaction strengths with growing electron count on Re, progressing from A = Cu, Ag, Na to Ca and further to La, Dy, Ce. This tendency is consistent with increasing Curie temperature which is observed experimentally, see $T^{\rm exp}_C$ in Table ~\ref{Tab:FeRe-exchanges}. Using the same data from Table ~\ref{Tab:FeRe-exchanges} one can calculate that each electron on Re contributes to the exchange energy of $\sim 100-130$ K/per bond depending on material. 

\begin{table}[t!]
\centering
\begin{tabular}{cccc}
\hline
\hline
 & $U^{\rm eff}_{\rm Cu}=6$~eV & $U^{\rm eff}_{\rm Cu}=7$~eV & $U^{\rm eff}_{\rm Cu}=8$~eV \\ \hline
 $J_{{\rm Fe - Re}}$, K & 72 & 73 & 83 \\
 $J_{{\rm Fe - Cu}}$, K & 18 & 24 & 18 \\
 $J_{{\rm Cu - Re}}$, K & 476 & 496 & 518 \\
 \hline
 $ T_C^{\rm MF}$, K & 1094 & 1122 & 1213  \\
$ T_C^{\rm ORF} $, K & 555 & 569 & 618 \\
\hline
\hline
 & $U^{\rm eff}_{\rm Fe}=3.1$~eV & $U^{\rm eff}_{\rm Fe}=4.1$~eV & $U^{\rm eff}_{\rm Fe}=5.1$~eV \\ \hline
 $J_{{\rm Fe - Re}}$, K & 81 & 73 & 81 \\
 $J_{{\rm Fe - Cu}}$, K & 14 & 24 & 18 \\
 $J_{{\rm Cu - Re}}$, K & 486 & 496 & 512 \\
 \hline
  $ T_C^{\rm MF}$, K & 1158 & 1122 & 1194  \\
$ T_C^{\rm ORF} $, K & 590 & 569 & 608 \\
\hline
\hline
 & $U^{\rm eff}_{\rm Re}=1$~eV & $U^{\rm eff}_{\rm Re}=1.5$~eV & $U^{\rm eff}_{\rm Re}=2$~eV \\ \hline
 $J_{{\rm Fe - Re}}$, K & 70 & 73 & 80 \\
 $J_{{\rm Fe - Cu}}$, K & 14 & 24 & 22 \\
 $J_{{\rm Cu - Re}}$, K & 456 & 496 & 532 \\\hline
  $ T_C^{\rm MF}$, K & 1057 & 1122 & 1217  \\
$ T_C^{\rm ORF} $, K & 538 & 569 & 618 \\
\hline\hline
\end{tabular}
\caption{\label{Tab:ExchangeMatrix} Results of DFT+U calculations for LaCu$_3$Fe$_2$Re$_2$O$_{12}$ with different $U^{\rm eff}=U-J_H$. In each case, $U^{\rm eff}$ for one of the elements was modified, while the other parameters were fixed at default values ($U^{\rm eff}_{\rm Cu}=7$~eV, $U^{\rm eff}_{\rm Fe}=4.1$~eV, $U^{\rm eff}_{\rm Re}=1.5$~eV)}. 

\end{table}

\subsection{Exchange mechanism}

One might use very different languages to explain the enhancement of AFM Cu-Re exchange with increasing electron number in Re $5d$ shell. In the localized electron picture, this could be interpreted by superexchange mechanism, where the growing number of pathways between half-filled Cu $x^2-y^2$ and Re $t_{2g}$ orbitals strengthens the interaction (the Re-O-Cu bond angle $\angle\mathrm{Re-O-Cu} \approx 110^\circ$ remains nearly constant across all materials).

However, taking advantage of the Green's function method, one can quantitatively evaluate orbital contributions to the total exchange interaction. This analysis reveals that there is a single large element between the Cu $x^2-y^2$ and Re $xy$ orbitals. Thus, in NaCu$_3$Fe$_2$Re$_2$O$_{12}$ the main contribution of 312~K due to Cu $x^2-y^2$ and Re $xy$ orbitals nearly completely determines the magnitude of the total exchange interaction (314~K) between these two atoms. It has to be also mentioned that the fact that $\angle\mathrm{Re-O-Cu}$  strongly deviates from $110^\circ$ makes overlap between $x^2-y^2$ and $xy$ orbitals possible.

\begin{figure}[b!]
\includegraphics[width=0.5\textwidth]{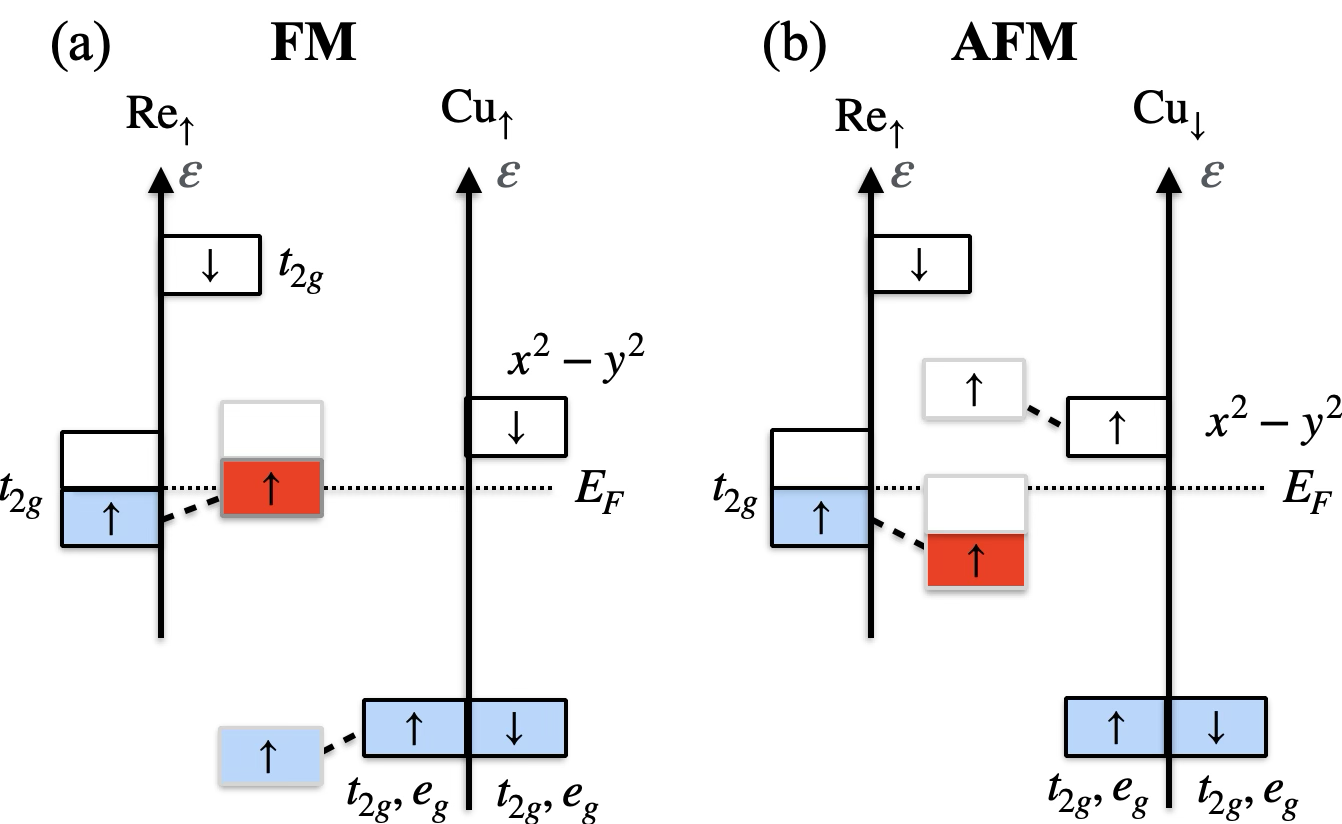}
\caption{\label{fig:Cu-Re-mechanism} Sketch of the electronic structure illustrating mechanism of the antiferromagnetic (AFM) exchange interaction between metallic Re $5d$ band crossing the Fermi level and more localized Cu $3d$ states, which are split due to strong Hubbard's repulsion in way that only the $x^2-y^2$ orbital remains half-filled. Hybridization (shown by dashed lines) is allowed only for the states of the same spins. In the case of FM order (a) it shifts Re $5d$ band upwards, while for (b) AFM order Re $5d$ states go lower. In the latter case the system gains the exchange energy, and corresponding coupling turns out to be AFM.}
\end{figure}

This suggests that description in terms of band magnetism can be more appropriate to analyze Cu - Re exchange interaction. In Fig.~\ref{fig:Cu-Re-mechanism} we compare two situations of FM and AFM orders between Cu and Re. Due to the Hubbard repulsion, the $x^2-y^2$ orbital is half-filled and appears slightly above the Fermi energy (at $\sim 1-2$ eV according to Fig.~\ref{fig:DOS}), while other occupied Cu $d$ states are deep below $E_F$ (at $\sim -5$ eV). The Re $t_{2g}$ states are exactly at the Fermi level. The hybridization, being possible only for the same spin states, pushes these Re $t_{2g}$ band upwards in the FM case  and increases the total energy. The situation is just the opposite for the AFM order: hybridization between half-filled Cu spin up $x^2-y^2$ and Re spin up $t_{2g}$ pulls the latter to lower energies and finally results in AFM exchange interaction between Re and Cu. Absolutely equivalent picture can be used to explain AFM coupling  for Re and Fe states.  We emphasize that this represents only a qualitative model, as quantitative estimates require detailed consideration of the specific band structure in each material.

The band mechanism of AFM coupling has several important implications. First, the energy gain is proportional to the band capacity, i.e., to the number of Re electrons. Furthermore, the same mechanism would contribute to FM coupling, if the $t_{2g}$ band were more than half-filled $ n_{{\rm Re-d}}^{{\rm nominal}}>3$. Therefore one might expect that maximal exchange is  close to the half-filling. This agrees with the results of direct DFT+U calculations, see Table~\ref{Tab:FeRe-exchanges}, and experimental observation of the largest Curie temperature for LaCu$_3$Fe$_2$Re$_2$O$_{12}$~\cite{liu2022} and DyCu$_3$Fe$_2$Re$_2$O$_{12}$~\cite{Liu2024} with $ n_{{\rm Re-d}}^{{\rm nominal}} = 2.5$. This also suggests that substituting Re with Os$^{4.5+}$, Ir$^{4+}$ or similar ions and maintaining  other constituents would weaken the AFM Cu-Re and Fe-Re exchange interactions, consequently reducing the Curie temperature.
\begin{figure}[t!]
\includegraphics[width=0.46\textwidth]{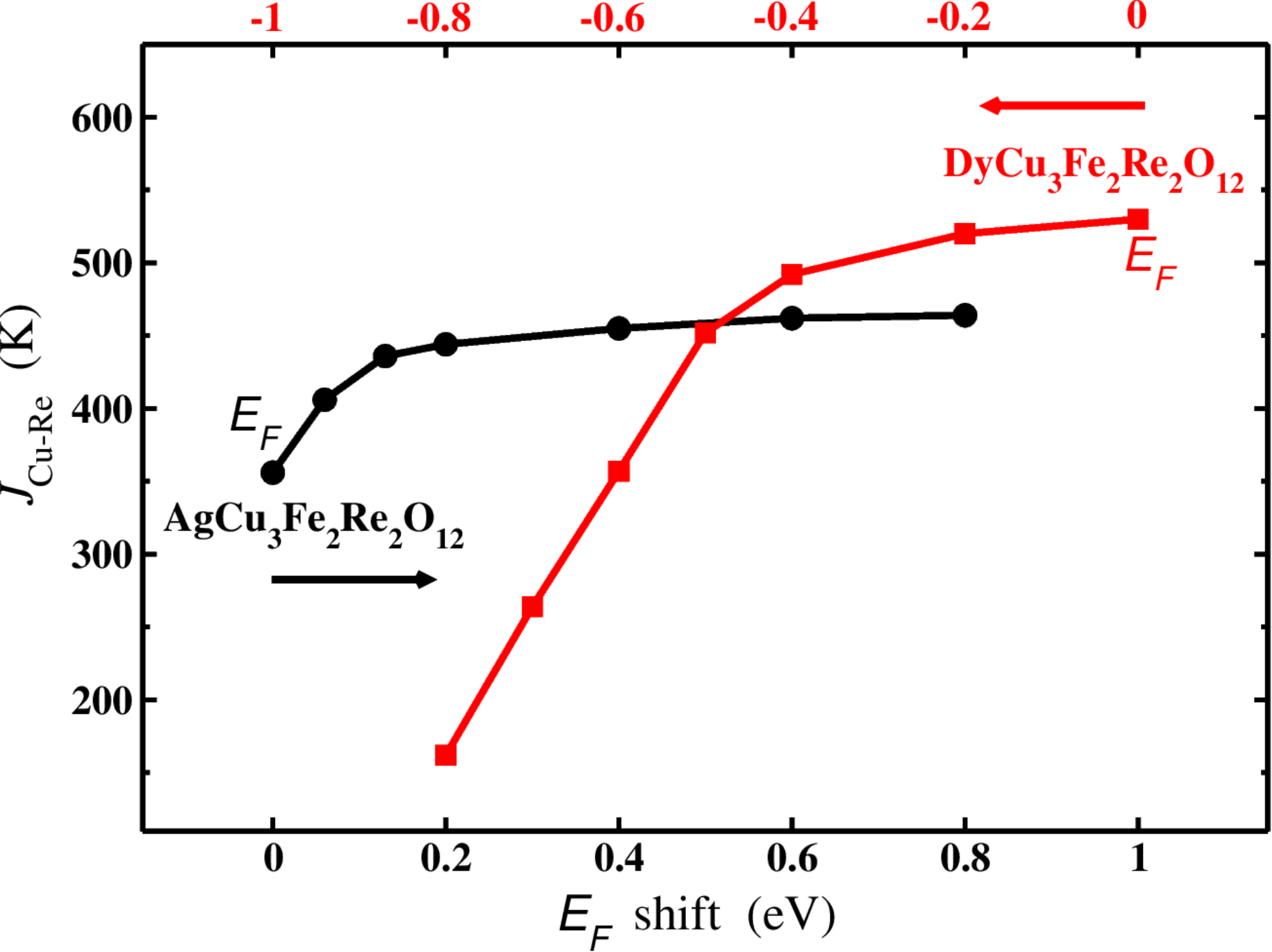}
\caption{\label{fig:exchangees-shift} Results of model calculations of Cu-Re exchange with variation of the Fermi level (Eq.~(11) in \cite{korotin2015}). The shift to low energies models decrease of number of Re $t_{2g}$ electrons in DyCu$_3$Fe$_2$Re$_2$O$_{12}$ (red), while increase of $E_F$ to higher energies simulates increase of Re $t_{2g}$ occupation for AgCu$_3$Fe$_2$Re$_2$O$_{12}$. Electronic structure was not recalculated.}
\end{figure}

In order to check the dependence on electron number  we performed model calculations artificially changing position of the Fermi level, when exchange constants are evaluated via the Green's function method. The results are presented in Fig.~\ref{fig:exchangees-shift}. One can clearly observe suppression of AFM Cu-Re exchange in DyCu$_3$Fe$_2$Re$_2$O$_{12}$ as we move $E_F$ downwards reducing number of Re $t_{2g}$ electrons. Interestingly, Fig.~\ref{fig:DOS} shows that the difference in position of the Fermi level between A=Dy and Ag is $\sim 0.6$ eV. Shifting $E_F$ by 0.6 eV (to lower energies) in DyCu$_3$Fe$_2$Re$_2$O$_{12}$ we obtain $J_{\rm Cu-Re} \sim 360$K, which is very close to results of direct calculations for AgCu$_3$Fe$_2$Re$_2$O$_{12}$. Moreover, the opposite tendency of increasing exchange interaction we observe starting from AgCu$_3$Fe$_2$Re$_2$O$_{12}$, where $ n_{{\rm Re-d}}^{{\rm nominal}}=1.5$ and now shifting $E_F$ upwards. Thus our model calculation clearly demonstrates strong dependence of the exchange interaction on the number of Re $t_{2g}$ electrons. Surprisingly, account of  this factor only is enough to explain suppression of exchange interaction (and consequently $T_C$) going from DyCu$_3$Fe$_2$Re$_2$O$_{12}$ to AgCu$_3$Fe$_2$Re$_2$O$_{12}$, but $J_{\rm Cu-Re}$ dependence on electron number of Re is not linear according to these calculations.  

Finally, we performed calculations for quadruple perovskite containing Os instead of Re:  CuCu$_3$Fe$_2$Os$_2$O$_{12}$. There are two types of Cu ions. Those (three Cu per formula unit) occupying CuO$_4$ plaquettes have 2+ valence state, while the one in icosahedral oxygen O$_{12}$ cage is 1+ similar to CuCu$_3$Fe$_2$Re$_2$O$_{12}$~\cite{pchelkina2023,poteryaev2025}. This gives 5.5+ oxidation state and $ n_{{\rm Os-d}}^{{\rm nominal}}=2.5$ for Os. $U-J_H=1.5$~eV for Os is the same as for Re. Calculation of exchange interaction leads to $J_{\rm Cu-Os} = 520$K and $J_{\rm Fe-Os} = 85$ K (suggesting $S=1$ for Os), which is very similar to that we had in LaCu$_3$Fe$_2$Re$_2$O$_{12}$ and DyCu$_3$Fe$_2$Re$_2$O$_{12}$ with $ n_{{\rm Re-d}}^{{\rm nominal}}=2.5$.

It is also worth mentioning that calculated exchanges do not strongly depend on choice of Hubbard repulsion (see Table~\ref{Tab:ExchangeMatrix}). Even increase of $U^{\mathrm{eff}} = U - J_H$ in two times changes $J_{\rm {Cu-Re}}$ only on $\sim 20$\%.

\section{Estimation of critical temperatures}

Having calculated the exchange constants, one can estimate the Curie temperatures. The simplest way to calculate magnetic properties is via the mean-field approach. For each sublattice $a$ the magnetization in the field $H$ per unit cell (containing $n_a$ atoms of the sort $a$) is given by: 
\begin{equation}
\label{Magnetization}
M_a=n_a\langle S^z_a \rangle=n_aS_a B_{S_a}\left(\frac{S_a H'_a}{T} \right),
\end{equation}
where 
\begin{equation}
\label{Hamiltonian}
H'_a=H-\sum_b z_{ab} J_{a b}M_b/n_b,
\end{equation}
 $B_S$ is the Brillouin function, 
 $z_{ab}$ is the number of nearest neighbors in the sublattice $b$ of an atom in the sublattice $a$. 
 Expanding for a small $H'$  and introducing $S_a(S_a+1)=\mu_a^2$ we
obtain
\begin{equation}
\label{Siz}
M_a =\chi_a^{(0)}(H-\sum_b z_{ab} J_{a b} M_b/n_b),
\end{equation}
where 
\begin{equation}
\label{chi}
\chi_a^{(0)}=n_a\mu_a^2/3T.
\end{equation}
Dividing by $H$ we obtain the system of equations for partial susceptibilities  of each sublattice, $\chi_a=M_a/H$,
\begin{equation}
\label{Chi}
\chi_a=\chi_a^{(0)}-\sum_b J_{a b}(0)\chi_a^{(0)}\chi_b/(n_a n_b)
\end{equation}
where  $J_{a b}(0) =J_{a b}({\bf q}= 0) =z_{a b}n_a J_{a b}$,  
 $z_{ab}n_a = z_{ba}n_b$ being the number of exchange-coupled pairs.

From the high-temperature expansion for the total susceptibility $\chi=\sum_a n_a\chi_a$, the paramagnetic  Curie-Weiss temperature is given by
\begin{equation}
\label{Chi1}
\theta=-\frac{1}{3}\sum_{a > b}J_{a b}(0)\mu_a^2\mu_b^2/\sum_{a} n_a\mu_a^2
\end{equation}
and is negative for a ferrimagnet with antiferromagnetic exchange interactions ($J_{a b}(0)>0$).

The condition for the  determinant 
\begin{equation}
|1 + J_{a b}(0)\chi_a^{(0)}/(n_a n_b)|=0
\end{equation}
with susceptibility $\chi_a^{(0)}$ depending on temperatures gives us the Curie temperature.
For quadruple perovskites we can use a three-sublattice model with exchange parameters between sublattices only, first sublattice corresponding to Fe ions, second to Re and third to Cu ones.  In this case one obtains the cubic equation for $T_C$ in the following standard form:
\begin{equation}
\label{cubic}
T_C^3+pT_C+q=0
\end{equation}
where
\begin{equation}
\begin{aligned}
&p=-\Big(\tilde J_{12}^2+\tilde J_{13}^2+\tilde J_{23}^2\Big),\\
&q=2\tilde J_{12}\tilde J_{13}\tilde J_{23}.
\end{aligned}
\end{equation}
and $\tilde J_{a b}=\frac{1}{3}\mu_a\mu_bJ_{a b}(0)/\sqrt {n_a n_b}$.
Since  the arithmetic mean  is greater than or equal to the geometric mean, we have
\begin{equation}
\label{Q}
Q=\Big(\frac{p}{3} \Big)^3+\Big(\frac{q}{2} \Big)^2 \leq 0.
\end{equation}
 Then the  solution of the cubic equation is obtained by the trigonometric substitution:
\begin{equation}
	T_C=2\sqrt {- \frac{p}{3}}\cos \left( \frac{1}{3}{\rm arccos}\left(\frac{3q}{2p}\sqrt\frac{-3}{p} \right) \right).
    \end{equation}
 If we neglect exchange interaction between Fe and Cu, $J_{13}=0$,  $T_C$ is given by
\begin{equation}
T_C=\sqrt{\tilde J_{12}^2+\tilde J_{23}^2} =\frac{1}{3}{\mu_2\sqrt{\frac{\mu_1^2 J_{12}^2(0)}{n_1 n_2}+\frac{\mu_3^2 J_{23}^2(0)}{n_2 n_3}}}. 
\label{t0}
\end{equation}

To take into account a possible partial itinerancy of electrons, the mean-field approach can be modified by generalizing the Heisenberg model to an effective interpolating model \cite{wysocki}. Then the moments $\mu_a$ may become non-integer being renormalized due to electron delocalization. 

To take into account fluctuation effects, we can use the Onsager reaction field (ORF) theory which is equivalent to the random-phase approximation and similar to the Tyablikov approach \cite{ORF}. For simplicity, we calculated the Onsager renormalization (see Table II) by taking into account two main exchange parameters only, $J_{12} = J_{\mathrm{Fe-Re}}$ and $J_{23} = J_{\mathrm{Re-Cu}}$, and putting $J_{13}=0$. Then the renormalized $T_C$ is given by ${T_C}={T_C^{(0)}}/I$ with the renormalization factor
\begin{equation}
I=\sum_{\bf q}\frac{ \tilde J_{12}^2(0)+\tilde J_{23}^2(0)}{ \tilde J_{12}^2(0)- \tilde J_{12}^2({\bf q})+ \tilde J_{23}^2(0)- \tilde J_{23}^2({\bf q})}.
\label{I}
\end{equation}
Here $T_C^{(0)}$ is the Curie temperature in the mean-field approximation, e.g., \eqref{t0}, and $q$-dependencies of exchange integrals for this particular lattice are 
\begin{eqnarray}
J_{12}({\bf q})&=&4 J_{12} \Big( 3+ \cos (\frac{q_x a}{2}) +\cos(\frac{q_y a}{2}) +\cos(\frac{q_z a}{2}) \Big),\nonumber \\
J_{23}({\bf q})&=&    J_{23} \Big(9 +3 \cos(\frac{q_x a}{4}+\frac{q_y a}{4}+\frac{q_z a}{4})  \nonumber \\
&+& 4\cos (\frac{q_x a}{4}-\frac{q_y a}{4}+\frac{q_z a}{4}) 
+ 4 \cos(\frac{q_x a}{4}+\frac{q_y a}{4}-\frac{q_z a}{4}) \nonumber \\
&+& 4 \cos(\frac{q_x a}{4}-\frac{q_y a}{4}-\frac{q_z a}{4}) \Big).
\end{eqnarray}

In Table~\ref{Tab:FeRe-exchanges} we present renormalized Curie temperatures for all studied materials. 
Interestingly, renormalization factor  varies only weakly, from $I=1.91$ to 1.98. The calculations demonstrate good overall agreement with the experimental Curie temperatures, further supporting the correctness of the exchange parameters obtained via the DFT+U method and the analysis performed using this approach.

\section{Conclusions}

In this paper, we investigated the electronic structure and magnetic interactions in the quadruple perovskites ACu$_3$Fe$_2$Re$_2$O$_{12}$ using DFT+U calculations and estimated Curie temperatures in the mean-field approach and Onsager reaction field theory. All the compounds were found to be half-metallic within  DFT+U due to Stoner-like splitting of delocalized Re 5$d$ states and strong on-site Hubbard repulsion of localized Fe and Cu $3d$ electrons. The magnetic structure is determined by strong antiferromagnetic exchange interactions between Re-Cu and Re-Fe which lead to the ferrimagnetic ground state. While  Cu and Fe moments are ferromagnetically aligned, conducting Re $t_{2g}$ states reside in the spin-minority channel.

Although all these compounds  have similar magnetic properties, there are notable differences. The A = Ca, La, Dy and Ce systems have high  $T_C$,  DFT+U calculations confirming large exchange parameter for these compounds. On the other hand, A = Cu, Ag and Na compounds have lower $T_C$, which is consistent with the smaller calculated exchange interactions. Besides that, the experimental value of magnetic moment for  CuCu$_3$Fe$_2$Re$_2$O$_{12}$ is strongly reduced. 
As demonstrated in Ref. \cite{poteryaev2025}, this may indicate violation of half-metallicity owing to strong correlation effects.

We propose a band mechanism to explain the nature and strength of Cu-Re and Fe-Re exchange interactions. Our calculations confirm partially itinerant features of rhenium  $d$-electrons suggested previously, a strong dependence of the exchange interactions on their number is demonstrated.  Calculations for A = Cu, Ag and Na yield considerably reduced Re magnetic moments.  The reduction in the number of $d$-electrons on Re, caused by substituting the A-site cation from A$^{3+}$ to A$^{2+}$ and A$^{1+}$, decreases the occupancy of the Re $t_{2g}$ states below the Fermi level. This, in turn, reduces the exchange interaction strength and consequently lowers $T_C$. Furthermore, model calculations demonstrate that shifting the Fermi level results a similar effect. So, the quadruple perovskites ACu$_3$Fe$_2$Re$_2$O$_{12}$ demonstrate both localized-spin (Fe and Cu) and itinerant features (Re) in magnetism. 

Theoretical values of $T_C$  based on calculated exchange parameters significantly overestimate the experimental $T_C$  values  in the mean-field approach. Tyablikov-Onsager renormalization provides a satisfactory agreement, but overestimates $T_C$ for A = Ag and Cu.  CuCu$_3$Fe$_2$Re$_2$O$_{12}$ stands out as the most exotic material among all studied quadruple perovskites with experimentally observed low saturation magnetization and anharmonic specific heat\cite{poteryaev2025}. Consequently, it is unsurprising that simple DFT+U calculations fail to achieve perfect agreement with experiment. This discrepancy arises from two main factors: first, the added complexity of rattling Cu ions in an icosahedral coordination~\cite{pchelkina2023}, and second, the significant role of many-body effects, as revealed by DMFT calculations. They result in a substantial reconstruction of the electronic structure, suppressing the half-metallic state \cite{poteryaev2025}, which in turn can renormalize the exchange interactions.

\section*{ACKNOWLEDGMENTS} 
The first-principle calculations were supported by the Russian Science Foundation via project RSF 23-42-00069, while estimation of the Curie temperature by Onsager reaction field theory was performed with support by Ministry of Science and Higher Education of the Russian Federation. Y. Long and Z. Liu were supported by the National Key R\&D Program of China (Grant No. 2021YFA1400300), and the National Natural Science Foundation of China (Grant Nos. 12425403, 12261131499, 12204516).

\bibliography{ref}

@unpublished{Long,
author = {Long, Y.},
note = {unpublished}
}

@article{Liu2024,
title = {High-pressure synthesis and high-performance half metallicity of quadruple perovskite oxide {DyCu$_3$Fe$_2$Re$_2$O$_{12}$}},
journal = {Fundamental Research},
year = {2024},
issn = {2667-3258},
url = {https://www.sciencedirect.com/science/article/pii/S2667325824005156},
author = {Zhehong Liu and Jinfeng Peng and Xiao Wang and Fedor Temnikov and Alexey Ushakov and Xubin Ye and Zhao Pan and Jie Zhang and Maocai Pi and Shuai Tang and Kai Chen and Florin Radu and Zhiwei Hu and Chien-Te Chen and Zhenhua Chi and Zlata Pchelkina and Valentin Irkhin and Yao Shen and Sergey V. Streltsov and Youwen Long}
}

@article{pchelkina2023,
  title = {Rattling Phonon Modes in Quadruple Perovskites},
  author = {Pchelkina, Z. V. and Komleva, E. V. and Irkhin, V. {\relax Yu}. and Long, Y. and Streltsov, S. V.},
  year = {2023},
  month = nov,
  journal = {JETP Letters},
  volume = {118},
  number = {10},
  pages = {738--741},
  issn = {0021-3640, 1090-6487},
  doi = {10.1134/S0021364023603202},
  urldate = {2025-04-24},
  copyright = {CC0 1.0 Universal Public Domain Dedication},
  langid = {english}
}

@article{liu2022,
  title = {Realization of a Half Metal with a Record-High Curie Temperature in Perovskite Oxides},
  author = {Liu, Zhehong and Zhang, Shuaikang and Wang, Xiao and Ye, Xubin and Qin, Shijun and Shen, Xudong and Lu, Dabiao and Dai, Jianhong and Cao, Yingying and Chen, Kai and Radu, Florin and Wu, Wen-Bin and Chen, Chien-Te and Francoual, Sonia and Mardegan, Jos{\'e} R. L. and Leupold, Olaf and Tjeng, Liu Hao and Hu, Zhiwei and Yang, Yi-feng and Long, Youwen},
  year = {2022},
  month = apr,
  journal = {Advanced Materials},
  volume = {34},
  number = {17},
  pages = {2200626},
  issn = {0935-9648, 1521-4095},
  doi = {10.1002/adma.202200626},
  urldate = {2025-04-14},
  langid = {english}
}

@article{poteryaev2025,
  title = {Highly Correlated Electronic State in the Ferrimagnetic Quadruple Perovskite {CuCu$_3$Fe$_2$Re$_2$O$_{12}$}},
  author = {Poteryaev, A. I. and Pchelkina, Z. V. and Streltsov, S. V. and Long, Y. and Irkhin, V. {\relax Yu}.},
  year = {2025},
  month = jan,
  journal = {JETP Letters},
  volume = {121},
  number = {1},
  pages = {67--71},
  issn = {0021-3640, 1090-6487},
  doi = {10.1134/S002136402460441X},
  urldate = {2025-04-23},
  copyright = {CC0 1.0 Universal Public Domain Dedication},
  langid = {english}
}

@article{perdew1996,
  title={Generalized gradient approximation made simple},
  author={Perdew, John P and Burke, Kieron and Ernzerhof, Matthias},
  journal={Physical review letters},
  volume={77},
  number={18},
  pages={3865},
  year={1996},
  publisher={APS}
}

@article{kresse1996,
  title={Efficient iterative schemes for ab initio total-energy calculations using a plane-wave basis set},
  author={Kresse, Georg and Furthm{\"u}ller, J{\"u}rgen},
  journal={Physical review B},
  volume={54},
  number={16},
  pages={11169},
  year={1996},
  publisher={APS}

}

@article{dudarev1998,
  title={Electron-energy-loss spectra and the structural stability of nickel oxide: {An LSDA+U} study},
  author={Dudarev, Sergei L and Botton, Gianluigi A and Savrasov, Sergey Y and Humphreys, CJ and Sutton, Adrian P},
  journal={Physical Review B},
  volume={57},
  number={3},
  pages={1505},
  year={1998},
  publisher={APS}
}

@article{CaCuCrReO,
  title={High-pressure synthesis of quadruple perovskite oxide {CaCu$_3$Cr$_2$Re$_2$O$_{12}$} with a high ferrimagnetic {Curie} temperature},
  author={Zhang, Jie and Liu, Zhehong and Ye, Xubin and Wang, Xiao and Lu, Dabiao and Zhao, Haoting and Pi, Maocai and Chen, Chien-Te and Chen, Jeng-Lung and Kuo, Chang-Yang and others},
  journal={Inorganic Chemistry},
  volume={63},
  number={7},
  pages={3499--3505},
  year={2024},
  publisher={ACS Publications}
}

@article{RCuMnO,
  title={Preparation under high pressures and neutron diffraction study of new ferromagnetic {RCu$_3$Mn$_4$O$_{12}$} ({R = Pr, Sm, Eu, Gd, Dy, Ho, Tm, Yb}) perovskites},
  author={S{\'a}nchez-Ben{\'\i}tez, J and Alonso, JA and Falc{\'o}n, H and Mart{\'\i}nez-Lope, MJ and De Andr{\'e}s, A and Fern{\'a}ndez-D{\'\i}az, MT},
  journal={Journal of Physics: Condensed Matter},
  volume={17},
  number={40},
  pages={S3063},
  year={2005},
  publisher={IOP Publishing}
}

@article{RCuMnO2,
  title={Enhancement of the {Curie} temperature along the perovskite series {RCu$_3$Mn$_4$O$_{12}$} driven by chemical pressure of {R}$^{3+}$ cations ({R} = rare earths)},
  author={Sanchez-Benitez, Javier and Alonso, Jos{\'e} Antonio and Mart{\'\i}nez-Lope, Mar{\'\i}a Jes{\'u}s and de Andres, Alicia and Fern{\'a}ndez-D{\'\i}az, Mar{\'\i}a Teresa},
  journal={Inorganic chemistry},
  volume={49},
  number={12},
  pages={5679--5685},
  year={2010},
  publisher={ACS Publications}
}

@article{magnetoresistance1,
  title={Large low-field magnetoresistance in perovskite-type {CaCu$_3$Mn$_4$O$_{12}$} without double exchange},
  author={Zeng, Z and Greenblatt, M and Subramanian, MA and Croft, M},
  journal={Physical review letters},
  volume={82},
  number={15},
  pages={3164},
  year={1999},
  publisher={APS}
}

@article{magnetoresistance2,
  title={Enhanced magnetoresistance in the complex perovskite {LaCu$_3$Mn$_4$O$_{12}$}},
  author={Alonso, JA and S{\'a}nchez-Ben{\i}tez, J and De Andr{\'e}s, A and Mart{\i}nez-Lope, MJ and Casais, MT and Mart{\i}nez, JL},
  journal={Applied physics letters},
  volume={83},
  number={13},
  pages={2623--2625},
  year={2003},
  publisher={American Institute of Physics}
}

@article{magnetoresistance3,
  title={Magnetoresistance and electronic structure of the half-metallic ferrimagnet {BiCu$_3$Mn$_4$O$_{12}$}},
  author={Takata, Kazuhide and Yamada, Ikuya and Azuma, Masaki and Takano, Mikio and Shimakawa, Yuichi},
  journal={Physical Review B—Condensed Matter and Materials Physics},
  volume={76},
  number={2},
  pages={024429},
  year={2007},
  publisher={APS}
}

@article{CaCuTiO,
  title={Giant dielectric constant response in a copper-titanate},
  author={Ramirez, AP and Subramanian, MA and Gardel, M and Blumberg, G and Li, D and Vogt, T and Shapiro, SM},
  journal={Solid state communications},
  volume={115},
  number={5},
  pages={217--220},
  year={2000},
  publisher={Elsevier}
}

@article{CaCuPiezo,
  title={Creation of Piezoelectricity in Quadruple Perovskite Oxides by Harnessing Cation Defects and Their Application in Piezo-Photocatalysis},
  author={Wang, Kai and Guo, Xiangyu and Han, Chen and Liu, Lihong and Wang, Zhiliang and Thomsen, Lars and Chen, Peng and Shao, Zongping and Wang, Xudong and Xie, Fang and others},
  journal={ACS nano},
  year={2025},
  publisher={ACS Publications}
}

@article{Zhang2025,
  title={{Large Manipulation of Ferrimagnetic Curie Temperature by A-Site Chemical Substitution in ACu$_3$Fe$_2$Re$_2$O$_{12}$ (A = Na, Ca, and La) Half Metals}},
  author={Zhang, Jie and Temnikov, Fedor and Ye, Xubin and Wang, Xiao and Pan, Zhao and Liu, Zhehong and Pi, Maocai and Tang, Shuai and Chen, Chien-Te and Pao, Chih-Wen and others},
  journal={Inorganic Chemistry},
  year={2025},
  publisher={ACS Publications}
}

@article{wang2021,
  title={Prediction of half-metallic ferrimagnetic quadruple perovskites {ACu$_3$Fe$_2$Re$_2$O$_{12}$} ({A = Ca, Sr, Ba, Pb, Sc, Y, La}) with high {C}urie temperatures},
  author={Wang, Duo and Shaikh, Monirul and Ghosh, Saurabh and Sanyal, Biplab},
  journal={Physical Review Materials},
  volume={5},
  number={5},
  pages={054405},
  year={2021},
  publisher={APS}
}

@article{Chen2014,
  title={{A half-metallic A-and B-site-ordered quadruple perovskite oxide CaCu$_3$Fe$_2$Re$_2$O$_{12}$ with large magnetization and a high transition temperature}},
  author={Chen, Wei-tin and Mizumaki, Masaichiro and Seki, Hayato and Senn, Mark S and Saito, Takashi and Kan, Daisuke and Attfield, J Paul and Shimakawa, Yuichi},
  journal={Nature Communications},
  volume={5},
  number={1},
  pages={3909},
  year={2014},
  publisher={Nature Publishing Group UK London}
}

@article{korotin2015,
  title={{Calculation of exchange constants of the Heisenberg model in plane-wave-based methods using the Green's function approach}},
  author={Korotin, Dm M and Mazurenko, VV and Anisimov, VI and Streltsov, SV},
  journal={Physical Review B},
  volume={91},
  number={22},
  pages={224405},
  year={2015},
  publisher={APS}
}

@article{ORF,
  title={Onsager reaction-field theory for magnetic models on diamond and hcp lattices},
  author={Wysin, GM},
  journal={Physical Review B},
  volume={62},
  number={5},
  pages={3251},
  year={2000},
  publisher={APS}
}

@article{wysocki,
  title={Thermodynamics of itinerant magnets in a classical spin-fluctuation model},
  author={Wysocki, Aleksander L and Glasbrenner, James K and Belashchenko, Kirill D},
  journal={Physical Review B—Condensed Matter and Materials Physics},
  volume={78},
  number={18},
  pages={184419},
  year={2008},
  publisher={APS}
}

@article{sun2022,
  title={Understanding the role of {Cu$^+$/Cu$^0$} sites at {Cu$_2$O} based catalysts in ethanol production from {CO$_2$} electroreduction - {A DFT} study},
  author={Sun, Liren and Han, Jinyu and Ge, Qingfeng and Zhu, Xinli and Wang, Hua},
  journal={RSC advances},
  volume={12},
  number={30},
  pages={19394--19401},
  year={2022},
  publisher={Royal Society of Chemistry}
}

@article{hong2012,
  title={Spin-phonon coupling effects in transition-metal perovskites: {A DFT+U} and hybrid-functional study},
  author={Hong, Jiawang and Stroppa, Alessandro and {\'I}niguez, Jorge and Picozzi, Silvia and Vanderbilt, David},
  journal={Physical Review B—Condensed Matter and Materials Physics},
  volume={85},
  number={5},
  pages={054417},
  year={2012},
  publisher={APS}
}

@article{scudder2021,
  title={Highly efficient transverse thermoelectric devices with {Re$_4$Si$_7$} crystals},
  author={Scudder, Michael R and He, Bin and Wang, Yaxian and Rai, Akash and Cahill, David G and Windl, Wolfgang and Heremans, Joseph P and Goldberger, Joshua E},
  journal={Energy \& Environmental Science},
  volume={14},
  number={7},
  pages={4009--4017},
  year={2021},
  publisher={Royal Society of Chemistry}
}

@article{kim2016,
  title={{Crystal structure and magnetism in $\alpha$-RuCl$_3$: An ab initio study}},
  author={Kim, Heung-Sik and Kee, Hae-Young},
  journal={Physical Review B},
  volume={93},
  number={15},
  pages={155143},
  year={2016},
  publisher={APS}
}

@article{lim2016,
  title={{Insights into cationic ordering in Re-based double perovskite oxides}},
  author={Lim, Tae-Won and Kim, Sung-Dae and Sung, Kil-Dong and Rhyim, Young-Mok and Jeen, Hyungjeen and Yun, Jondo and Kim, Kwang-Ho and Song, Ki-Myung and Lee, Seongsu and Chung, Sung-Yoon and others},
  journal={Scientific reports},
  volume={6},
  number={1},
  pages={19746},
  year={2016},
  publisher={Nature Publishing Group UK London}
}

\end{document}